# Predicting Tau Accumulation in Cerebral Cortex with Multivariate MRI Morphometry Measurements, Sparse Coding, and Correntropy

Jianfeng Wu[a], Wenhui Zhu[a], Yi Su[b], Jie Gui[c], Natasha Lepore[d], Eric M. Reiman[b], Richard J. Caselli[e], Paul M. Thompson[f], Kewei Chen[b], Yalin Wang[a*]

[a] School of Computing, Informatics, and Decision Systems Engineering, Arizona State University, Tempe, USA; [b] Banner Alzheimer's Institute, Phoenix, USA; [c] School of Cyber Science and Engineering, Southeast University, Nanjing, China, [d] CIBORG Lab, Department of Radiology Children's Hospital Los Angeles, Los Angeles, USA, [e] Department of Neurology, Mayo Clinic Arizona, Scottsdale, USA; [f] Imaging Genetics Center, Stevens Neuroimaging and Informatics Institute, University of Southern California, Marina del Rey, USA

## ABSTRACT

Biomarker-assisted diagnosis and intervention in Alzheimer's disease (AD) may be the key to prevention breakthroughs. One of the hallmarks of AD is the accumulation of tau plaques in the human brain. However, current methods to detect tau pathology are either invasive (lumbar puncture) or quite costly and not widely available (Tau PET). In our previous work, structural MRI-based hippocampal multivariate morphometry statistics (MMS) showed superior performance as an effective neurodegenerative biomarker for preclinical AD and Patch Analysis-based Surface Correntropy-induced Sparse coding and max-pooling (PASCS-MP) has excellent ability to generate low-dimensional representations with strong statistical power for brain amyloid prediction. In this work, we apply this framework together with ridge regression models to predict Tau deposition in Braak12 and Braak34 brain regions separately. We evaluate our framework on 925 subjects from the Alzheimer's Disease Neuroimaging Initiative (ADNI). Each subject has one pair consisting of a PET image and MRI scan which were collected at about the same times. Experimental results suggest that the representations from our MMS and PASCS-MP have stronger predictive power and their predicted Braak12 and Braak34 are closer to the real values compared to the measures derived from other approaches such as hippocampal surface area and volume, and shape morphometry features based on spherical harmonics (SPHARM).

## 1. INTRODUCTION

Alzheimer's disease (AD) is now viewed as a gradual process that begins many years before the onset of clinical symptoms. Measuring brain biomarkers and intervening at the preclinical or earlier stages of AD are believed to improve the probability of therapeutic success[1–3]. Amyloid-β (Aβ) deposition and Tau tangles are the two specific protein pathological hallmarks of Alzheimer's disease (AD) and play crucial roles in leading to dementia related structural deformations observed in magnetic resonance imaging (MRI) scans[4–7]. In the A/T/N system - a recently proposed research framework for understanding the biology of AD - the presence of abnormal levels of Tau in the brain or cerebrospinal fluid (CSF) is used to define the presence of biological Alzheimer's disease[2]. An imbalance between production and clearance of Aβ occurs early in AD and is typically followed by the accumulation of tau protein tangles with increasingly severe atrophy and neurodegeneration detectable on brain MRI scans[2,3,8]. In particular, the hippocampus is a primary target region across the spectrum from clinically normal to dementia[9–12]. Cognitive unimpaired individuals with abnormal Aβ burden have faster progression of hippocampal volume atrophy[13,14]. Tau burden has a strong correlation with subsequent hippocampal volume atrophy[7]. Tau pathology can be measured using positron emission tomography (PET) with Tau-sensitive radiotracers or in CSF. Even so, PET scans are not as widely available as MRI. The high cost and invasiveness of CSF and PET measurements make them less attractive to subjects in the preclinical AD stage, leading to interests in less invasive predictors of Tau burden from MRI. Even a moderately accurate predictor of Tau burden could be a valuable precursor prior to more invasive testing.

* Send correspondence to Yalin Wang
 E-mail: ylwang@asu.edu

In our recent work[15,16], we used hippocampal multivariate morphometry statistics (MMS) together with a sparse coding algorithm, Patch Analysis-based Surface Correntropy-induced Sparse coding and max-pooling (PASCS-MP), to predict Aβ positivity. In this work, we further leverage a similar framework to predict two measurements of Tau deposition, especially in brain areas closely related to the Braak staging and named as Braak12 and Braak34[17–20]. As shown in **Figure 1**, the framework consists of two parts. First, from MRI scans, we extract MMS for the hippocampus in both brain hemispheres. The MMS are high-fidelity vertex-wise surface morphometry features, based on using multivariate tensor-based morphometry (mTBM) to encode morphometry along the surface tangent plane and radial distance (RD) to encode morphometry along the surface normal direction. They have been proven to outperform more traditional measures in detecting clinically relevant group differences[21–23]. However, the high dimensional features are not suitable for machine learning models, especially when the sample size is small. Therefore, we introduce an unsupervised feature extraction method, PASCS-MP, to generate a low-dimensional representation for each subject. In PASCS-MP, we first randomly select patches of MMS features across the hippocampal surface. Then, the correntropy-induced sparse coding model is applied to generate sparse codes for these patches, which can reduce the negative influence of non-Gaussian noise in the MMS. Finally, the max-pooling method is leveraged to reduce the dimensionality of these sparse code and a low-dimensional representation is generated for each subject. With these representations, we then train ridge regression model

### *(1) Extract Multivariate Morphometry Statistics from MR images*

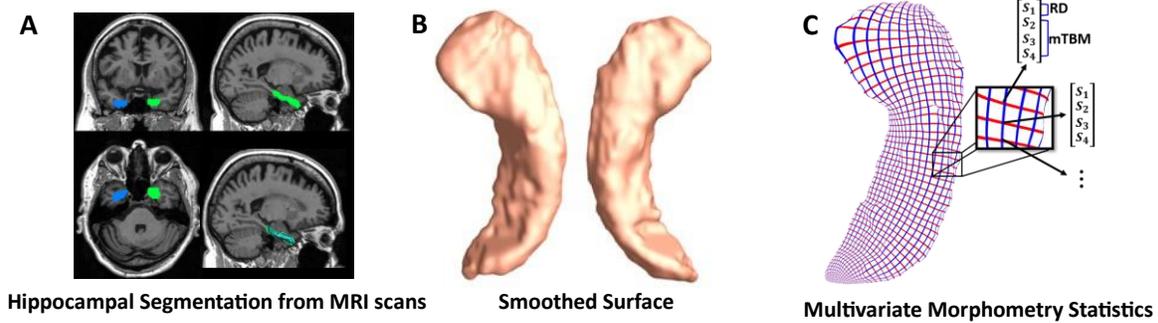

**Hippocampal Segmentation from MRI scans**    **Smoothed Surface**    **Multivariate Morphometry Statistics**

### *(2) Our proposed machine learning system involving PASCS-MP and Ridge regression to predict Tau measurements*

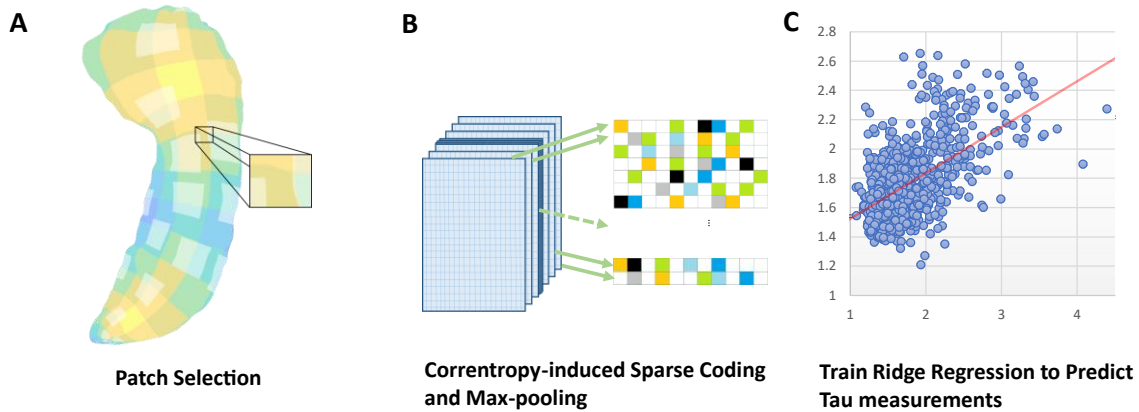

**Patch Selection**    **Correntropy-induced Sparse Coding and Max-pooling**    **Train Ridge Regression to Predict Tau measurements**

**Figure 1. Framework Overview**. Panel (1) shows hippocampal Multivariate Morphometry Statistics (MMS) are extracted from MR images. MMS is a $4 \times 1$ vector on each vertex, including radial distance (scalar) and multivariate tensor-based morphometry ($3 \times 1$ vector). Panel (2) shows our Patch Analysis-based Surface Correntropy-induced Sparse-coding (PASCS) method. In subfigure A, we randomly select patches of MMS one the hippocampal surface. Then, the Correntropy-induced Sparse-coding and max-pooling methods are used to generate a low-dimensional representation for each subject. In subfigure C, ridge regression models are trained with these representations to predict Braak12 and Braak34 measurements. (The data in this subfigure is synthetic) Finally, the prediction model is validated with a 10-fold cross-validation scheme.

to predict Braak12 and Braak34 measurements. We hypothesize that our MMS-based PASCS-MP may provide stronger predictive power than the traditional hippocampal volume, surface area, and spherical harmonics (SPHARM) based hippocampal shape measurements.

## 2. DATA DESCRIPTION

Data for testing the performance of our proposed framework and comparable methods were obtained from the publicly available ADNI database[24] (adni.loni.usc.edu). ADNI is the result of efforts by many co-investigators from a broad range of academic institutions and private corporations. Subjects are recruited from over 50 sites across the U.S. and Canada. From ADNI 1, ADNI 2, ADNI GO, and ADNI 3 (the different phases of ADNI), we obtained 925 pairs of MRI scans and AV1451 PET images. The PET images are reprocessed using a single pipeline consistent with the work of Sanchez et al.[25], so the standardized uptake value ratio (SUVR) from different ADNI study sites can be analyzed together. In this work, we analyze tau deposition in two brain regions named Braak12 and Braak34[17–20]. **Table 1** shows the demographic information from the cohort that we analyzed.

Table 1. Demographic information for the subjects we study from the ADNI.

| Cohort | Group | Sex (M/F) | Age | MMSE | Braak12 | Braak34 |
|---|---|---|---|---|---|---|
| ADNI (n=925) | AD (n=115) | 62/53 | 76.0±8.5 | 22.0±4.5 | 2.39±0.60 | 2.51±0.73 |
| | MCI (n=278) | 158/120 | 74.6±7.9 | 27.9±2.1 | 1.82±0.46 | 1.92±0.46 |
| | CU (n=532) | 210/322 | 73.4±7.1 | 29.1±1.1 | 1.58±0.23 | 1.73±0.21 |

Values are mean ± standard deviation where applicable.

## 3. METHODS

### 3.1 Surface Multivariate Morphometry Statistics

In this work, we extract a high-fidelity vertex-wise surface morphometry feature, multivariate morphometry statistics, from each MRI scan. We first use FIRST (FMRIB's Integrated Registration and Segmentation Tool)[26] to segment hippocampus substructures from the 3D volumetric brain MRI images. A topology-preserving level set method[27] and marching cubes algorithm[28] are used to construct triangular surface meshes for each pair of hippocampi. The surface is further smoothed with mesh simplification using *progressive meshes*[29] and mesh refinement by the Loop subdivision surface method[30]. Our prior studies[31–38] have shown that the smoothed meshes are accurate approximations to the original surfaces, with a higher signal-to-noise ratio.

To facilitate hippocampal shape analysis, we generate a conformal grid ($150 \times 100$) on each surface, which is then used as a canonical space for surface registration. On each hippocampal surface, we compute its conformal grid with a holomorphic 1-form basis[38,39]. We adopt surface conformal representation[35,36] to obtain surface geometric features for automatic surface registration. In our system, we further extend the surface fluid registration method to an inverse-consistent framework [40]. The obtained surface registration is diffeomorphic. For details of our inverse-consistent surface fluid registration method, we refer to the paper of Shi et al.[36]. Each surface has the same number of vertices ($150 \times 100$) as shown in panel 2 of **Figure 1**. The intersection of the red curve and the blue curve is a surface vertex, and at each vertex, we adopt two types of features (with a total of 4 dimensions), the radial distance (RD) and the surface metric tensor used in multivariate tensor-based morphometry (mTBM). The RD (a scalar at each vertex) represents the thickness of the shape at each vertex to the medical axis[41,42]. this reflects the surface differences along the surface normal directions. The mTBM statistics (a vector at each vertex) have been frequently studied in our prior work[36,38,43]; they measure local surface deformation along the surface tangent plane. The surface of the hippocampus in each brain hemisphere has 15,000 vertices, so the feature dimensionality for the hippocampus in each subject is 60,000.

### 3.2 PASCS-MP

MMS are vertex-wise surface morphometry features, describing the regional atrophy or expansion[44] and development[45] of the surface. However, the dimensionality of MMS is much higher than the number of subjects, which leads to the so-called *high dimension-small sample problem.* Also, outliers and redundant information in MMS will also affect the prediction accuracy for linear regression models. Feature reduction methods proposed by the work[46,47] may ignore the intrinsic properties of a structure's regional morphometry. Therefore, we introduce the following feature reduction method[48] for the vertex-wise surface morphometry features.

To extract useful surface features and reduce the dimension before making predictions, this work first randomly generates square windows on each surface to obtain a collection of small image patches with different amounts of overlap as shown in the zoomed-in window in subfigure A of panel (2) in **Figure 1.** After that, we extract meaningful features with sparse coding and dictionary learning[49]. Dictionary learning has been successful in many image processing tasks as it can concisely model natural image patches. In this work, we leverage a novel sparse coding and dictionary learning method with an $l_1$-regularized correntropy loss function named *Correntropy-induced Sparse-coding (CS)*, which is expected to improve the computational efficiency compared to Stochastic Coordinate Coding (SCC)[50]. Formally speaking, correntropy is a generalized similarity measure between two scalar random variables U and V, which is defined by $\mathcal{V}_\sigma(U,V) = \mathbb{E}\mathcal{K}_\sigma(U,V)$. Here, $\mathcal{K}_\sigma$ is a Gaussian kernel given by $\mathcal{K}_\sigma(U,V) = \exp\{-(u-v)^2/\sigma^2\}$ with the scale parameter $\sigma > 0$, $(u-v)$ being a realization of $(U,V)$[51,52]. Using the correntropy measure as a loss function will reduce the negative influence of non-Gaussian noise in the data.

Classical dictionary learning techniques[53,54] consider a finite training set of feature maps, $X = (x_1, x_2, \ldots, x_n)$ in $R^{p \times n}$. In our study, $X$ is the set of MMS features from $n$ surface patches of all the samples. All the MMS features on each surface patch, $x_i$, is reshaped to a $p$-dimensional vector. We desire to generate a new set of sparse codes, $Z = (z_1, z_2, \ldots, z_n)$ in $R^{m \times n}$ for these features. Therefore, we aim to optimize the empirical cost function in **Eq. (1)**:

$$f(D, z_i) \triangleq \sum_{i=1}^n l(x_i, D, z_i) \quad (1)$$

where $D \in R^{p \times m}$ is the dictionary and $z_i \in R^m$ is the sparse code of each feature vector. $l(x_i, D, z_i)$ is the loss function that measures how well the dictionary $D$ and the sparse code $z_i$ can represent the feature vector $x_i$. Then, $x_i$ can be approximated by $x_i = Dz_i$. In this way, we convert the $p$-dimensional feature vector, $x_i$, to a $m$-dimensional sparse code, $z_i$, where $m$ is the dimensionality of the sparse code (the dimensionality can be arbitrary). In this work, we introduce the correntropy measure[52] to the loss function and define the $l_1$-sparse coding optimization problem in **Eq. (2):**

$$\min_{D, z_i} \{1 - \sum_{i=1}^n \exp\left(-\frac{\|Dz_i - x_i\|_2^2}{\sigma^2}\right) + \lambda \sum_{i=1}^n \|z_i\|_1\} \quad (2)$$

where $\lambda$ is the regularization parameter, $\sigma$ is the kernel size that controls all properties of correntropy. $\|\cdot\|_2$ and $\|\cdot\|_1$ are the $l_2$-norm and $l_1$-norm and *exp()* represents the exponential function. The first part of the loss function measures the degree of the image patches' goodness and the correntropy may help remove outliers. Meanwhile, the second part is well known as the $l_1$ penalty[55] that can yield a sparse solution for $z_i$ and select robust and informative features. Specifically, there are $m$ columns (atoms) in the dictionary $D$ and each atom is $d_j \in R^p, j = 1, 2, \ldots, m$. To prevent $D$ from being arbitrarily large and leading to arbitrary scaling of the sparse codes, we constrain each $l_2$-norm of each atom in the dictionary to be no larger than one. We will let $C$ become the convex set of matrices verifying the constraint in **Eq. (3)**:

$$C \triangleq \{D \in R^{p \times m} s.t. \forall j = 1, 2, \ldots, m, d_j^T d_j \leq 1\} \quad (3)$$

Note that, the empirical problem cost $f(D, z_i)$ is not convex when we jointly consider the dictionary $D$ and the coefficients $Z$. But the function is convex concerning each of the two variables, $D$, and $Z$, when the other one is fixed. As it takes much time to solve $D$ and $Z$ when dealing with large-scale data sets and a large-size dictionary, we adopt the framework in the stochastic coordinate coding (SCC) algorithm[50], which can dramatically reduce the computational cost of the sparse coding, while keeping a comparable performance.

After we get the sparse code (the dimension is *m*) for each patch, the dimensionality of sparse codes for each subject is still too large for classification, i.e., $m \times 1008$. Therefore, we apply max-pooling process to reduce the feature dimensionality for each subject. Max-pooling[56] is a way of taking the most responsive node of a given region of interest and serves as an important layer in the convolutional neural network architecture. In this work, we compute the maximum value of a particular feature over all the sparse codes of a subject and generate a new representation for each subject, which is an *m*-dimensional vector. These summary representations are much lower in dimension, compared to using all the extracted surface patch features; this can improve the generalizability of results via less over-fitting. Finally, these representations are used as features to train ridge regression models to predict Tau measurements.

## 4. EXPERIMENTAL RESULTS

**4.1 Key Parameter Estimations for the PASCS-MP Method**

Before applying the PASCS-MP method to the MMS, four key parameters need to be selected empirically, including the patch size, the dimensionality of the learned sparse coding, the regularization parameter for the $l_1$-norm (λ) and the kernel size (σ) in the exponential function. Rational parameters will benefit to refining AD-related MMS representations. Instead of predicting Tau measurements, we train ridge regression models to predict MMSE on a separate dataset from ADNI (100 AD patients, 100 MCI, and 100 CN). We perform grid search to explore the optimal parameter settings. After performing 10-fold cross-validation ten times, we compare the average root mean squared errors (RMSE) of MMSE for each parameter setting. In **Figure 2**, we only illustrate the average and 95% confidence interval of RMSE for part of the grid search result. In each subfigure, we only compare one parameter and fix the remaining three. Eventually, we find that the optimal patch size is 10×10, the optimal sparse code dimensionality is 1800, the optimal λ is 0.22, and the optimal σ is 3.6; these optimal parameters are subsequently adopted for predicting Tau measurements.

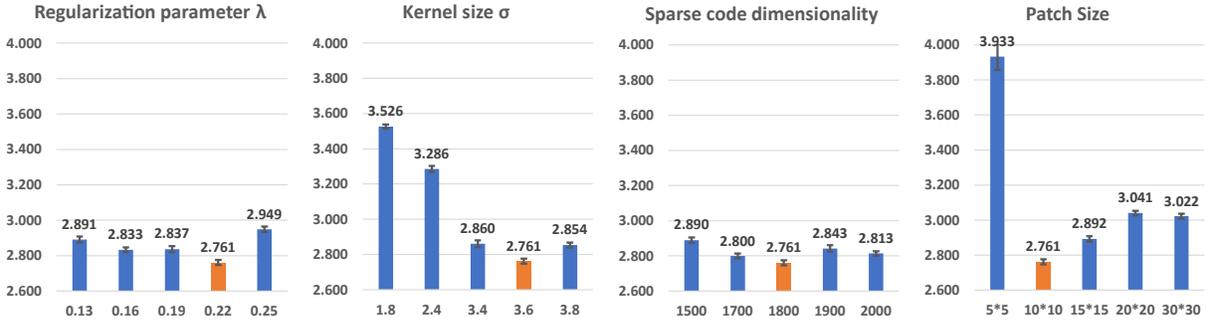

**Figure 2.** The relationship of each parameter to RMSE. The *x*-axis represents the value for each parameter. The orange bars represent the classification performances using the optimal parameters. Each bar represents the average and 95% confidence interval for RMSE.

**4.2 Prediction of Tau Measurements**

After performing PASCS-MP on MMS of 925 subjects from ADNI, we obtain 925 new representations, of which the dimensionality is 1,800. These representations are utilized for training ridge regression models to predict two Tau measurements, Braak12 and Braak34. For each measurement, we also repeat the 10-fold cross-validation ten times. The mean and 95% confidence interval of the RMSE for the two measurements are illustrated in **Figure 3**. To demonstrate that our representations have stronger predictive power, we train ridge regression models with hippocampal surface area, hippocampal volume, and the hippocampal shape features calculated by the popular SPHARM method[36,57]. As shown in **Figure 3**, our PASCS-MP always has the minimum RMSE.

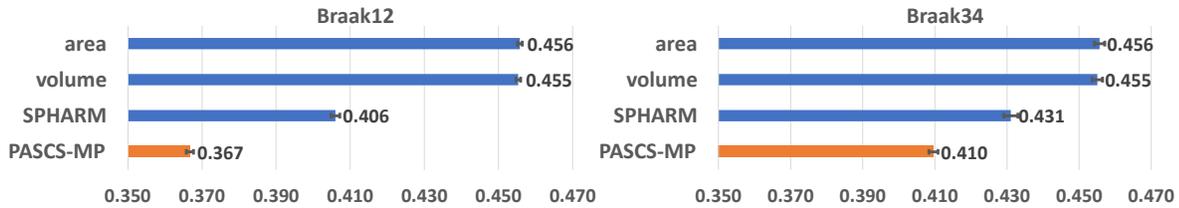

**Figure 3.** RMSE for predicted Braak12 and Braak34 from four measurements, hippocampal surface area, volume, SPHARM and our MMS-based PASCS-MP representations. Each bar represents the mean and 95% confidence interval of RMSE for ten 10-fold cross-validations.

**4.3 Analysis of the Predicted Tau Measurements**

To evaluate the predicted Tau measurements from different features, we first perform analysis of variance (ANOVA) among the three clinical groups, AD, MCI, and CU. The distributions of the predicted Tau measurements are shown in **Figure 4**. The first column is the distribution of real Braak12 and Braak34. Other columns are the predicted Tau measurements from hippocampal surface area, hippocampal volume, SPHARM, and our PASCS-MP. The *F*-value and *p*-value of ANOVA among the three clinical groups are illustrated in each subfigure. Our PASCS-MP achieves the most significant group difference among all the predicted Tau measurements.

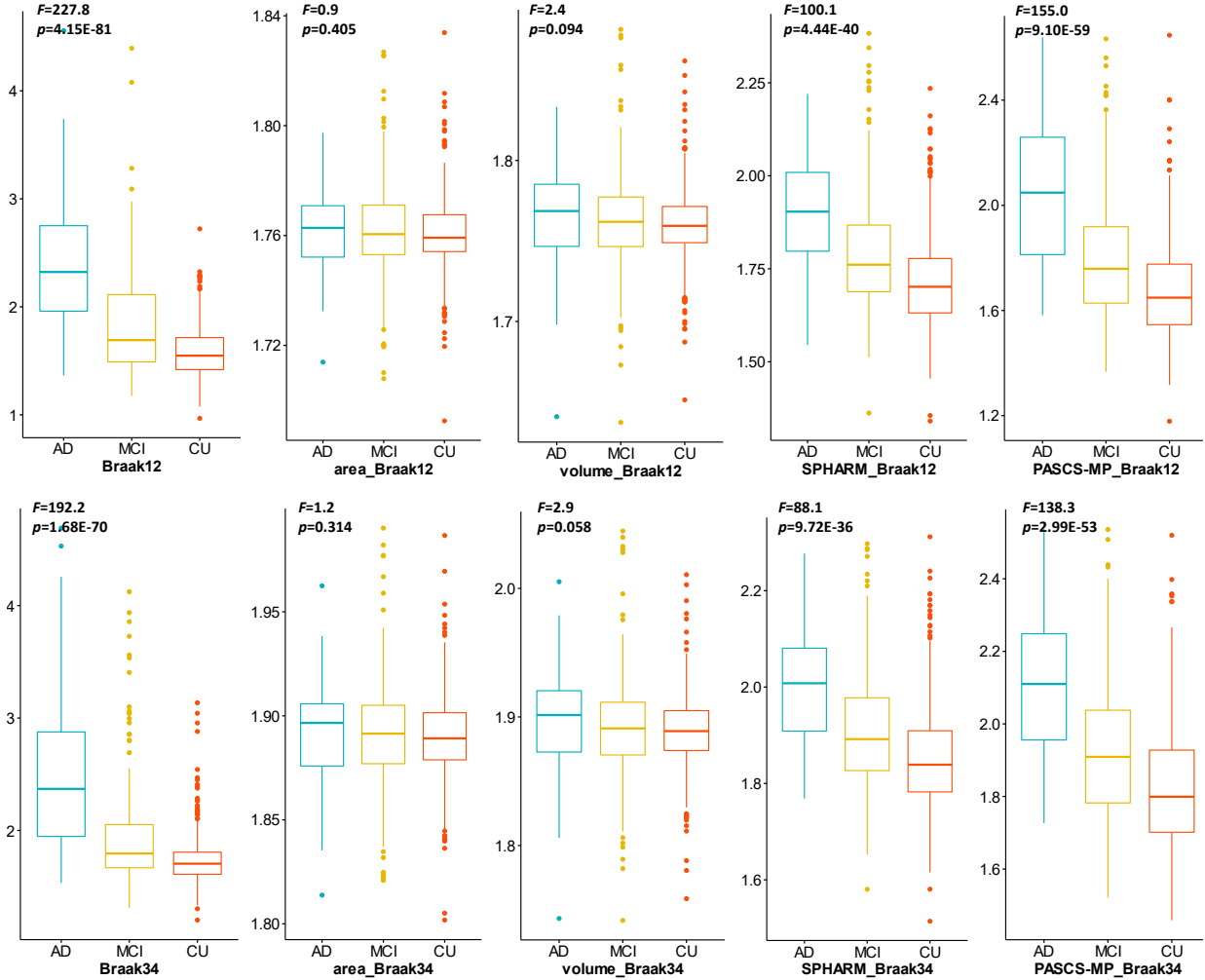

**Figure 4.** The first column is the distribution of real Braak12 and Braak34 measures. The remaining columns are the predicted Tau measurements from hippocampal surface area, hippocampal volume, SPHARM, and our MMS-based PASCS-MP representations. The *F*-value and *p*-value of ANOVA among AD, MCI, and CU are illustrated on the top of each subfigure.

In addition, we leverage the Pearson correlation to evaluate the relations between real Tau measurements and each of the predicted Tau measurements. In **Figure 5**, we visualize the linear relationships. The vertical axis is the real Tau measurement, and the horizontal axis is the predicted one. The correlation coefficient, *R*, and *p*-value for each analysis are also illustrated in each subfigure. Our PASCS-MP always has the largest correlation coefficients, in these experiments, compared to the traditional measurements, which means the Tau measurements predicted by our MMS-based PASCS-MP representations are close to the real Tau measurements. Both experiments demonstrate that our MMS-based PASCS-MP representations have the best accuracy of the approaches we examined for predicting Tau measurements.

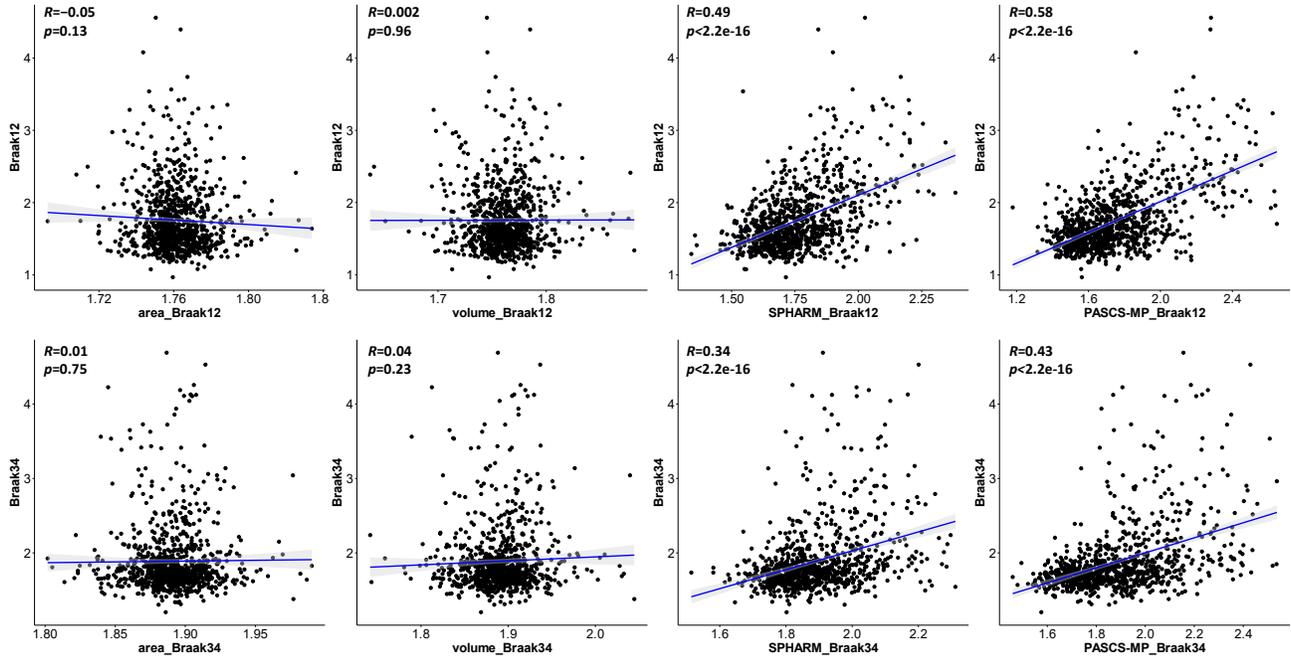

**Figure 5.** The first row shows the Pearson correlation between real Braak12 and predicted Braak12 from hippocampal surface area, hippocampal volume, SPHARM, and our MMS-based PASCS-MP representations. The second row shows the Pearson correlation between real Braak34 and predicted Braak34. The *y*-axis is the real Tau measurement and *x*-axis shows the predicted Tau measurement. The Pearson correlation coefficient, *R*, and *p*-values are in the top left corner of each subfigure.

## 5. CONCLUSION

In this paper, we explore the association between hippocampal structures and Tau deposition in 925 subjects from ADNI. Compared to the traditional hippocampal shape measurements, our MMS-based representations refined by PASCS-MP achieve better performance in predicting the measurements of Tau deposition. The resulting prediction has a smaller root mean squared errors than those predictions obtained with other features such as hippocampal surface area, hippocampal volume, and SPHARM. The correlation between the predicted tau accumulation values and the real values is also better for the presented methodology, and an ANOVA test shows significant differences in the predicted values for the different patient groups. In the future, we will use this framework to study other AD-related regions of interest (ROIs) and further improve the framework to visualize the disease-related features on the surface.